# MATERIALS RESEARCH FOR HIPER LASER FUSION FACILITIES: CHAMBER WALL, STRUCTURAL MATERIAL AND FINAL OPTICS

J. Alvarez, A. Rivera, R. González-Arrabal, D. Garoz, E. del Río, J.M. Perlado

*Instituto de Fusión Nuclear (UPM), José Gutiérrez Abascal 2, E28006 Madrid, Spain*
antonio.rivera.mena@gmail.com

*The European HiPER project aims to demonstrate commercial viability of inertial fusion energy within the following two decades. This goal requires an extensive Research & Development program on materials for different applications (e.g., first wall, structural components and final optics). In this paper we will discuss our activities in the framework of HiPER to develop materials studies for the different areas of interest. The chamber first wall will have to withstand explosions of at least 100 MJ at a repetition rate of 5-10 Hz. If direct drive targets are used, a dry wall chamber operated in vacuum is preferable. In this situation the major threat for the wall stems from ions. For reasonably low chamber radius (5-10 m) new materials based on W and C are being investigated, e.g., engineered surfaces and nanostructured materials. Structural materials will be subject to high fluxes of neutrons leading to deleterious effects, such as, swelling. Low activation advanced steels as well as new nanostructured materials are being investigated. The final optics lenses will not survive the extreme ion irradiation pulses originated in the explosions. Therefore, mitigation strategies are being investigated. In addition, efforts are being carried out in understanding optimized conditions to minimize the loss of optical properties by neutron and gamma irradiation.*

## I. INTRODUCTION

The European inertial fusion project, HiPER (High Power laser Energy Research facility) is now in the final year of the preparatory phase (phase 2). The goal is to build a facility for repetitive laser operation (HiPER 4a) working with bunches of 100 shots with up to 5 consecutive ignition shots and a maximum energy per bunch of about 100 MJ. Next, (circa 5 years later) a reactor to demonstrate commercial viability of inertial fusion energy (IFE) will be built (HiPER 4b). Thus, it will have to operate continuously at full power producing its own tritium and generating electric power. Table I offers an overview of the currently planned operation scenarios for the HiPER 4a and 4b phases. Over the next 7 years, a technological phase (HiPER phase 3) will take place to minimize construction risks through appropriate R&D activities. Currently, there are several potential solutions for chamber first wall, structural materials, final optics and reactor systems that will be downselected based on R&D activities undertaken by HiPER and other inertial fusion projects, remarkably, NIF and LIFE programs in the US[1].

One of our main tasks is the development of advanced materials with well defined properties. A common feature is radiation resistance. In general, the use of self-healing materials is an advantage[2]. Radiation induced defects do not accumulate in these materials. In particular, they show a reduced accumulation of vacancies (precursors of cavities and extended defects). This can be achieved by enhancing vacancy recombination. For example, the mean free path of vacancies in nanostructured materials is comparable to the grain size. Thus, vacancy annihilation at grain boundaries is greatly enhanced.

TABLE I. Operation scenarios currently planned for HiPER construction phases 4a and 4b.

|  | HiPER 4a | HiPER 4b |
|---|---|---|
| Description | Experimental facility | Demonstration reactor |
| Operation | Bunches of 100 shots, max. 5 DT explosions | Continuous (24/7) |
| Target yield (MJ) | <20 | >100 |
| Rep. rate (Hz) | 1-10 | 10-20 |
| Thermal power (GW) | - | 1-3 |
| Tritium cycle | No | Yes |
| Blanket | No | Yes |

## II. YIELD FLUXES AND CHAMBER CHOICE

The ignition scheme is not decided yet but it is desirable to keep open several options: central ignition with direct and indirect targets as well as advanced high gain schemes based on direct targets such as fast ignition and shock ignition[3,4].

The ignition scheme and especially the target type (i.e., direct or indirect) strongly affect the choice of chamber type. Table II shows the yields produced by both types of targets. The most interesting feature is that the fraction of energy released as X-rays amounts to 25% for indirect targets whereas for direct targets about 27% of the released energy is in the form of ions. The prompt release of X-rays with indirect targets makes unavoidable some kind of protection strategy for the chamber walls. Well developed projects such as LIFE propose a residual gas (typically a few μg/cm$^3$ of Xe) although other possibilities such as wet walls exist. Note that gas protection is incompatible with the propagation of a PW, ps laser needed for fast ignition schemes and probably with the injection and tracking of direct targets. In order to keep on considering these schemes we will not consider for the moment the use of gas protection strategies but the use of evacuated dry wall chambers. The rest of the paper discusses our activities on **materials** studies for **evacuated dry chambers.**

TABLE II. Typical yields produced by direct and indirect targets[5]. The figures indicate percentage of total energy released from the target in one explosion.

|  | Direct target | Indirect target |
|---|---|---|
| Neutrons | 71% | 69% |
| X-rays | 1% | 25% |
| Pulse width | a few ns | a few ns |
| Avg. energy | 10 keV | 1 keV |
| Ions | 27% | 6% |
| Burn products | 13% | 2% |
| Pulse width | 200 ns |  |
| Avg. energy | 2 MeV |  |
| Debris ions | 14% | 4% |
| Pulse width | 1.5 μs |  |
| Avg. energy | 150 keV |  |

### III. FIRST WALL MATERIALS

In the case of **direct targets** the energy deposited to the chamber walls in the form of ions arrives delayed due to the ion time of flight (see the pulse widths in Table II). For this reason the thermal load is considerably lower than in the case of indirect targets of similar yield. This opens the possibility to **use evacuated dry walls,** provided that appropriate materials are available. This is certainly the case for the **HiPER phase 4a** scenario based on low yield targets. A mm-sized W armor is enough to protect the chamber structural material under this scenario. A previous study by the authors[6] show that 50 MJ targets produce a relatively low increase in the W temperature (1200 K) and an acceptable mechanical deformation. Only thousands of shots might cause an important fatigue and cracking or relevant ion-driven damage but that will not be the case for HiPER4a, which is meant to withstand just a few hundred of energetic explosions.

The situation is very different under the strict operation conditions of **HiPER 4b** with target yields exceeding 100 MJ at high repetition rate in a continuous mode (1-3·10$^6$ shots/day). In this case no first wall material appears appropriate to withstand the fusion events at reasonable distances of about 5–6 m. This is due to thermo-mechanical failures such as cracking and melting and ion-driven damage, e.g., exfoliation of W through He nucleation in radiation-induced cavities[7,8]. In addition, little is known about synergistic effects related to the simultaneous implantation of H, D, T, He and C ions at high fluxes.

Thus we have launched a R&D program on materials for the development of first wall materials under the following requirements: (i) large surface area to accommodate the thermal load over a larger volume; (ii) high thermal conductivity to impede excessive heating due to reduced thermal removal; (iii) porous materials to facilitate the release of He and other light species; (iv) self-healing materials i.e. nanocrystals in which vacancies easily migrate to grain boundaries reducing the formation of large vacancy clusters and thus He nucleation. In this direction, there are works on the design of velvets, dendrites, foams and micro/nano structured W[9-12].

Our activities are based on two major types of **materials: (i) based on W and (ii) based on C.** Within the first group we are growing nanocrystalline W by sputtering. The goal is to get stable nano-grains under the specific irradiation conditions of inertial fusion. In addition we are modifying high quality W developed by the magnetic fusion community for the divertor region. The goal is to enhance the surface area by micro-engineering techniques in such a way that high thermal loads can be accommodated.

The second group of materials includes carbon based materials such as nanotubes, (nano)diamond and Ti doped carbides with very interesting properties such as high sublimation point and thermal conductivity. Known problems have been reported in magnetic fusion reactors mainly related to swelling, tritium retention and chemical sputtering. The self-healing properties of these materials are expected to prevent swelling. In addition, the conditions in inertial fusion, in particular the spherical geometry of the reaction chamber and the possibility of higher working temperatures might reduce tritium retention and chemical sputtering. Moreover, other strategies not suitable in magnetic fusion reactors could be employed as for example periodic annealing of the first wall at very high temperature.

The number of facilities to mimic real inertial fusion conditions is much lower than the number of analogous facilities for plasma-wall interaction studies relevant for the magnetic fusion community. It is remarkable the work developed for first wall studies within the framework of the American projects Aries, **NIF**, HAPL and LIFE[1,9,13,14]. It is of particular interest the unique facility **RHEPP** (Sandia National Laboratories) for studies of high ion fluxes as those obtained with direct targets[9]. Nowadays, it is the only facility able to qualify materials for thousands of cycles under irradiations that mimic conditions obtained with direct targets. All our samples designed for first wall applications will be studied in RHEPP as a first step prior to final qualification in a future dedicated facility.

It is important to stress that the mentioned investigations, in particular those focused on the handling of high heat loads are also relevant to the magnetic fusion community since, from a thermo-mechanical point of view, none of the so-far studied materials can withstand the most disadvantageous magnetic fusion conditions, e.g. disruptions[15].

**IV. FINAL OPTICS**

A very important issue for the development of inertial fusion is the final optics because the ignition process itself relies on its reliability to locate the laser beam energy on the target repeatedly at frequencies as high as 10-20 Hz. As yet, HiPER considers for final optics transmission lenses located at 8 m from the target in order to have the focusing precision required for direct targets. The optical material of choice for HiPER is silica due to its resistance to radiation degradation[16,17].

According to our calculations, at this distance **fused silica** reaches the melting temperature with direct targets of 48 MJ due to the energy deposited by ions in the first few micron thickness of the lenses. Mitigation is needed to preserve the optics in good conditions. While gas protection is ruled out due to incompatibilities with fast ignition schemes and direct target injection another option is to use electric and magnetic fields to avoid that the ions reach the fused silica. Note that since the lenses must be located at only 8 m from the target, electric or magnetic diversion require very high fields in each beam line. In case electric diversion is chosen, up to 800 kV will be necessary over distances of a few meters for full ion mitigation. Therefore, ion mitigation appears necessary but far from trivial. In a full mitigation scenario, only neutrons, gammas and X-rays reach the final optics. The energy deposited by the X-ray (target of 48 MJ) leads to a temperature rise of 700ºC at the surface of the sample and induce MPa stress due to the thermo-mechanical response. The value of the stress is below the safety limits but the repetitive cycle must be studied to avoid failures in points of contact with cooling surfaces.

Neutron and gamma radiation modify the properties of fused silica by means of atomistic effects: (i) the creation of color centers reduce the transmittance and (ii) densification takes place modifying the refraction index and thus the focal length of the lenses. These effects need special attention since they might be fatal for the operation of the reactor. One can in principle minimize or avoid this loss of efficiency, if the final optics components are kept at elevated temperatures (about 500ºC) enough to promote defect annealing[18]. However, careful experiments are needed to corroborate these expectations under realistic inertial fusion conditions.

For the final optics, we will compare **high quality optical graded silica samples with KU1 silica**, well known for its radiation degradation resistance[17]. It is important to mention that optical lenses for laser applications will always be coated by an anti-reflectance layer that will receive an important radiation dose. Therefore, we will pay attention to possible threatening effects on the coating layers (not well considered so far).

We are carrying out detailed calculations to estimate the lifetime of the silica components subjected to the combined effect of neutron and gamma irradiation. Note that contrary to the materials for first wall applications, the final optics performance is not well understood even under the HiPER 4a scenario. Taking into account the high neutron fluxes reaching the samples we have already proposed to carry out experiments at NIF to study the combined effect of neutrons and gammas on the samples.

**V. STRUCTURAL MATERIALS**

Structural materials will be subjected to severe in-service conditions such as high levels of radiation damage, high temperatures and coolants effects in the advanced nuclear energy systems. The extreme conditions planned for advanced nuclear reactors will affect material properties and their behavior under these aggressive environments. So, a proper selection of the structural materials, able to support these conditions, is very important in order to ensure the safe operation and design of all future nuclear installations[19,20].

**High-Cr ferritic/martensitic steels** are the leading candidate structural materials for key components in most future nuclear energy options[21]. Their high resistance to radiation effects such as swelling and damage accumulation[22-25] added to the better resistance against corrosion for high chromium contents[26] are the main reasons for this. Nevertheless, these alloys present problems of irradiation embrittlement. This effect could be caused by defects created by the irradiation as they could act as obstacles for the motion of dislocations. Therefore, the mechanical response of these materials will depend on the type of defects created during irradiation. Experiments have shown that the concentration and type of defects observed depend on Cr concentration, among other factors. Although the addition of Cr to the steels improves their properties against radiation damage, this improvement presents a non-monotonic trend of radiation hardening, embrittlement or swelling as a function of Cr concentration. Understanding of this effect is needed.

Experiments in conditions as close as possible to those expected in operation are necessary. For this purpose efforts are being carried out worldwide. However, these experiments are very expensive in terms of time and resources and the real operating conditions cannot be fully reproduced. For this reason, atomic level studies are important tools to study the response to irradiation of FeCr alloys and the multiscale modeling is the more extended way of performing these studies. Density functional theory (DFT) is currently the most accurate methodology to perform this type of calculations. But its high computational cost makes this method only available for a few hundreds of atoms. In order to increase time and length scales it is necessary to employ appropriate empirical interatomic potentials specially developed for these systems and properly describing their behavior under irradiation.

A new version of the concentration-dependent model potential (CDM) for FeCr compounds has been developed by A. Caro. Originally this potential[27] was adjusted to the heat of formation of the FeCr solid solution and was used to derive thermodynamic properties of the solution[28-30]. This new version was fitted to the main features of point defects in Fe-Cr[28]. We have tested the reliability of this new version of the CDM potential in radiation damage studies performing calculations of formation energies for a large variety of defects in both bcc Fe, Cr and FeCr solutions. We have compared the results with those obtained with DFT calculations by Olsson[31] and with another empirical potential specially developed for FeCr alloys, the two band-model[32].

We have also performed a detailed study of the dependence of the vacancy formation energy on Cr concentration[32]. We have performed calculations not only on the formation energy of the vacancy as a function of the Cr content for concentrations ranging from 1 to 17% Cr, but also of the relative position of the Cr atoms with respect to the vacancy. We have used for this study the new version of the concentration-dependent model potential mentioned above[32]. Currently we are studying the effect of the Cr concentration on the formation and binding energies of vacancy clusters up to 5 vacancies. We intend to extend these studies to other defects such as self and mixed interstitials.

## VI. CONCLUSIONS

The HiPER project will soon enter in the Technological phase with the objective of carrying out intensive R&D activities to minimize the risks of building a demonstration reactor for inertial fusion. Two scenarios are proposed for the construction phase: HiPER 4a, a experimental facility to check the validity of the technology and HiPER 4b, a reactor with full capabilities for demonstration purposes.

Currently, HiPER intends to keep open different ignition schemes based on both direct and indirect targets. Downselection will be done along the HiPER technological phase. Therefore, we are studying materials for different applications in the most versatile reaction chamber, i.e., an evacuated dry wall chamber.

There exist materials appropriate for the first wall under HiPER 4a conditions. Similarly, appropriate structural materials can be found. The situation is different with respect to the final optics (as yet based on transmission lenses). Ongoing work is being devoted to establish the lifetime of the final optics components in these conditions both, by means of calculations and experiments.

The situation under the strict HiPER 4b conditions is very different. Right now, no material can be used for the first wall due to fatal failure. An ambitious program in materials research is being carried out to find appropriate candidates, based either on W or on C. The performance of silica as a final optics component requires further efforts. One of the major difficulties is to carry out experiments able to mimic the high neutron fluxes typical of inertial fusion. Finally, a serious computational effort is devoted to study structural materials by means of multiscale modeling with a special focus on high Cr-ferritic-martensitic steels.

## ACKNOWLEDGMENTS

The authors want to thank the Spanish Ministry of Science and Innovation for the ACI-PROMOCIONA support 2009. This work is being carried out with the support of the Fusion Experimental Facilities of the project Technofusion (Madrid).